# LONG-LIVED PLASMA FORMATIONS IN THE ATMOSPHERE AS AN ALTERNATIVE ENERGY SOURCE

**M. S. Dvornikov,[1,2]* G. Sh. Mekhdieva,[3]† and L. A. Agamalieva[3]‡**

[1] N. V. Pushkov Institute of Terrestrial Magnetism, Ionosphere, and Radio Wave Propagation, Moscow, Russia
[2] National Research Tomsk State University, Tomsk, Russia
[3] Baku State University, Baku, Republic of Azerbaijan

*A model of a stable plasma formation, based on radial quantum oscillations of charged particles, is discussed. The given plasmoid is described with the help of the nonlinear Schrödinger equation. A new phenomenon of the effective attraction between oscillating charged particles is considered within the framework of the proposed model. The possible existence of a composite plasma structure is also discussed. Hypothesis about using the obtained results to describe natural long-lived plasma formations, which can serve as alternative energy sources, is put forward.*

## INTRODUCTION

A model of a plasma formation based on spherically or axially symmetric quantum oscillations in plasma is developed. In Section 1, on the basis of a nonlinear Schrödinger equation describing the radial motion of electrons, the dispersion relation is obtained for the given oscillations. The characteristic length scale of the plasmoid, which lies in the nanoscale range, is also obtained [1]. Next, in Section 2, we consider the effective interaction between oscillating charged particles inside the plasma structure due to the exchange of a virtual acoustic wave [2]. It has been determined that in the case of a dense plasma such an interaction can be attractive and lead to the formation of bound states of oscillating particles [3]. Within the framework of the proposed model, the possible existence of a composite plasma formation is discussed, in which multiple kernels are present, each of them is a nanoscale oscillation of the plasma (Section 3). The coupling between individual plasmoids is due to a quantum exchange interaction between ions [4]. Finally, in Section 4, we discuss the application of the obtained results to a theoretical description of long-lived plasma structures observed in nature.

## 1. QUANTUM RADIAL OSCILLATIONS OF ELECTRONS

We shall now lay out a model of a quantum plasma formation based on the nonlinear Schrödinger equation. The given result is published in [1].

The evolution of an electron gas in plasma obeys the following nonlinear Schrödinger equation [5]:

$$i\hbar \frac{\partial \psi}{\partial t} = \hat{H}\psi,$$

$$\hat{H} = -\frac{\hbar^2}{2m}\nabla^2 + e\varphi(\boldsymbol{r},t) + e^2 \int d^3 \boldsymbol{r}' \frac{|\psi(\boldsymbol{r}',t)|^2}{|\boldsymbol{r}-\boldsymbol{r}'|},$$

(1)

---

* E-mail: maxdvo@izmiran.ru
† E-mail: m.gulnara@hotmail.com
‡ E-mail: ag.leyla@hotmail.com

where $\psi(r,t)$ is the wave function, normalized to the electron gas density, $n_e = |\psi(r,t)|^2$, $\varphi(r,t)$ is the potential of the electrostatic interaction between the ions and the electrons, $m$ is the mass of the electron, and $e$ is the charge of the proton.

An approximate solution of Eq. (1) corresponding to spherically symmetric oscillations of the plasma was found in [1] in the form $\psi(r,t) = \psi_0 + \chi(r)e^{-i\omega t}$, where $|\psi_0|^2 = n_0$ is the unperturbed density and $\chi \sim \sin\gamma r / r$ is a small perturbation of the wave function. The frequency of the quantum oscillations $\omega$ depends on the parameter $\gamma$ as follows:

$$\gamma^2 = \frac{\omega m}{\hbar}\left[1 \pm \left(1 - 4\frac{\omega_e^2}{\omega}\right)^{1/2}\right]. \tag{2}$$

Here $\omega_e = \sqrt{4\pi n_0 e^2 / m}$ is the Langmuir frequency for electrons.

The characteristic radius of the plasma formation described by Eq. (1), which corresponds to $\omega = 2\omega_e$, can be found using formula (2). The result is $R = \pi(\hbar / 2m\omega_p)^{1/2}$. Assuming that we have singly ionized plasma with particle number density $n_0 = 2.7 \cdot 10^{19}$ cm$^{-3}$, we find that the radius of the plasmoid lies in the nanoscale range: $R = 1.4 \cdot 10^{-7}$ cm.

## 2. EFFECTIVE ATTRACTION BETWEEN CHARGED PARTICLES IN A QUANTUM PLASMOID

We will show that charged particles inside a quantum plasmoid can form bound states. This result was published in [2, 3]. In Section 1, it was shown that it is possible to create a spherically symmetric plasma structure based on quantum radial oscillations of electrons. Taking into account the fact that the radius of such a plasmoid lies in the nanoscale range, it can be expected that inside this object various quantum phenomena will occur.

Let us consider the case of plasma with a relatively low temperature. In this case, in addition to electrons and ions, neutral atoms will also be present. In this situation, rapid oscillations of charged particles will excite acoustic waves in the neutral component. If such an acoustic wave is coherently absorbed by another charged particle, this will lead to a non-Coulomb effective interaction between the charged particles. We will show that such an effective interaction can take place in a dense plasma. Moreover, in some situations it can be attractive.

Note that the process of the exchange of virtual acoustic waves between ions in dense plasma was investigated for the first time in [6], where it was asserted that this form of an effective interaction is important for the stability of the plasmoid. The given effective interaction was considered in [2]. With the help of a qualitative analysis, it was shown in [2] that the exchange of a virtual acoustic wave can lead to the formation of a bound state between two charged particles inside a plasmoid of nanoscale size.

Note that the idea that charged particles in a plasma can form bound states due to various quantum effects was considered earlier in a number of works (see, for example, [7]). It should also be noted that the attraction due to quantum effects differs from the well-known effective interaction between charged particles with the same polarity in a dusty plasma [8].

A discussion of a quantum effective interaction of charged particles inside a plasmoid should begin with the construction of the ground state of the system, i.e., the state with a minimal energy. A ground state based on particle wave functions corresponding to plane waves is unsuitable for further applications since it does not completely reflect the spherical symmetry of the system. In [3], it was proposed to use ground state wave functions corresponding to a three-dimensional harmonic oscillator with the frequency $\omega$, where $\omega$ is the frequency of the ion acoustic wave. In this case, the wave function in the momentum representation has the form,

$$\psi_{n\sigma}(p) = 2\pi \left[ \frac{n!}{\Gamma(l'+n+3/2)} \right] \left( \frac{\hbar}{m_i \omega_i} \right)^{3/4} \left( \frac{p^2}{m_i \omega_i \hbar} \right)^{l'/2}$$
$$\times \exp\left[ -\frac{p^2}{2 m_i \omega_i \hbar} \right] L_n^{l'+1/2}\left( \frac{p^2}{m_i \omega_i \hbar} \right) \chi_\sigma, \tag{3}$$

where $n$ and $\sigma$ are the radial and spin quantum numbers, $m_i$ is the mass of an ion, $\omega_i$ is the Langmuir frequency for ions, $L_n^\alpha(z)$ are the generalized Laguerre polynomials, $l'$ is the effective orbital quantum number, and $\chi_\sigma$ is the spin wave function.

The following Hamiltonian in the second quantization representation describes the interaction between charged particles and the acoustic wave field:

$$\hat{H}_{int} = K_0 \int d^3 r \hat{\psi}^\dagger r \hat{\psi}(r) \hat{n}_1(r). \tag{4}$$

Here $\hat{\psi}(r)$ is the wave function of the charged particles in the second quantization representation, which can be obtained from formula (3), $\hat{n}_1(r)$ is the perturbation of the acoustic field, and $K_0$ is a phenomenological interaction constant.

After the standard elimination of acoustic degrees of freedom, the full Hamiltonian takes the form,

$$\hat{H} = \sum_{n\sigma} e_n \hat{a}^\dagger_{n\sigma} \hat{a}_{n\sigma} - \sum_{nn'\sigma} F_{nn'} \hat{a}^\dagger_{n\sigma} \hat{a}^\dagger_{n',-\sigma} \hat{a}_{n',-\sigma} \hat{a}_{n\sigma}. \tag{5}$$

Here $\hat{a}^\dagger_{n\sigma}$ and $\hat{a}_{n\sigma}$ are creation and annihilation operators of the oscillatory perturbations of the charged particles, $e_n = E_n - \mu$, where $E_n$ is the energy of an oscillating charged particle, $\mu$ is the chemical potential of the system, and

$$F_{nn'} = \frac{4 K_0^2 n_n^{(0)} m_i}{3\pi^2 \hbar \omega_i m_n} \left( \frac{m_i \omega_i}{\hbar} \right)^{3/2} \frac{\tilde{n}^{3/2}}{\sqrt{nn'}} \left( 1 + \frac{3\hbar \omega_i}{16 m_i c_s^2 \tilde{n}} \right), \tag{6}$$

where $n_n^{(0)}$ is the number density of unperturbed neutral atoms, $m_n$ is the mass of a neutral atom, $c_s$ is the sound speed, and $\tilde{n} = \min(n, n')$.

In [3], it was shown that the Hamiltonian in formulas (5) and (6) describes the effective attraction between the charged particles in plasma, which, in turn, leads to the formation of bound states of oscillating particles. Employing relation (6), we can obtain the characteristic radius of a plasmoid, inside which such a pairing takes place:

$$R_{cr} = \frac{4 \hbar n_n^{(0)} \sigma_s}{3\pi m_n \omega_i}. \tag{7}$$

Here $\sigma_s$ is the total cross section of scattering of the ions off neutral atoms. If we assume that a spherical plasma formation based on oscillations of protons has arisen in neutron matter with particle number density $n_n^{(0)} = 10^{38}$ cm$^{-3}$, which corresponds to matter in a compact star, then we find, using formula (7), that the radius of the plasmoid has the value $R_{cr} = 4.47 \cdot 10^{-10}$ cm.

Note that, in the derivation of this estimate for the radius of the plasmoid, we have assumed that only pairs of ions (protons), rather than electrons, form pairs. This assumption is owing to the fact that the Langmuir frequency for protons is significantly less than that for electrons. Moreover, the scattering cross section for protons off neutral particles, i.e., off neutrons, is significantly greater than for electrons.

Despite the fact that the used density of matter and the obtained radius of the plasmoid lie beyond the limits of the values that can be expected under terrestrial conditions, the hypothesis was advanced in [3] that under certain conditions pairing of charged particles due to the exchange of a virtual acoustic wave can be implemented inside a spherically symmetric structure existing in an atmospheric plasma.

## 3. COMPOSITE PLASMA STRUCTURES

Let us describe the formation of a composite plasma structure within the framework of the proposed model. The given result was published in [4].

Let us discuss the possibility of the emergence of a composite plasma object consisting of individual kernels, at each of them a spherically symmetric plasma oscillation has been excited. The attraction between the plasmoids is due to the quantum exchange interaction between ions, which exist inside a plasma structure of nanoscale size. Note that the importance of the exchange interaction for a description of the stability of spherically symmetric plasma structures was noted earlier in [9].

We assume that the ions are particles with spin 1/2. The Hamiltonian corresponding to the exchange interaction between such ions has the form,

$$H_{ex} = -\sum_{i \neq j} J_{ij} \left( \hat{\mathbf{S}}_i \cdot \hat{\mathbf{S}}_j \right), \tag{8}$$

where $J_{ij}$ is the exchange integral and $\hat{\mathbf{S}}_i$ is the ion spin operator. The sum in formula (8) runs over pairs of ions located in different plasmoids.

Employing relation (8), it is possible to obtain the total energy of the exchange interaction of two oscillating plasma structures $V_{ex}$. It is convenient to normalize the given quantity by the energy of the electromagnetic interaction in the plasmoid $W_{em}$. Finally, the expression takes the form,

$$\frac{V_{ex}}{W_{em}} = -6.0 \cdot 10^{-3} \cdot J_0 \sigma^3 \cos \Delta\varphi \left( \frac{p_0 n_0}{T} \right)^2 F(a,b),$$

$$F(a,b) = b \exp\left( -\frac{a^2}{4} \right) \left\{ \frac{1+4b^2}{a} \left[ \operatorname{erfcx}\left( 2b - \frac{a}{2} \right) - \operatorname{erfcx}\left( 2b + \frac{a}{2} \right) \right] - b \left[ \operatorname{erfcx}\left( 2b - \frac{a}{2} \right) + \operatorname{erfcx}\left( 2b + \frac{a}{2} \right) \right] \right\}. \tag{9}$$

Here $J_0$ is a positive constant characterizing the magnitude of the exchange interaction, $\sigma$ is the effective radius of the plasmoid, $\Delta\varphi$ is the phase shift between oscillations in the two plasmoids, $p_0$ is the electric dipole moment of the ion, $n_0$ is the number density of the ions, $T$ is the temperature of the ions, $\operatorname{erfcx}(z)$ is the normalized error function, $a = R/\sigma$ and $b = \kappa\sigma$, $R$ is the distance between the plasmoid centers, $\kappa = \sqrt{2ME_i}/\hbar$, $M$ is the mass of an ion, and $E_i$ is the average energy of an ion in one of the two plasmoids.

To find a numerical estimate, let us consider a pair of plasmoids arising in a water plasma with the number density $n_0 \sim 10^{23} \text{ cm}^{-3}$ and the temperature $T \sim 300 \text{ K}$. Employing relation (9), we find that the quantum exchange interaction between plasmoids can be attractive for a plasma structure with a size lying in the nanoscale range:

$\sigma \sim 10^{-7}$ cm. In some situations, the ratio in formula (9) can reach several percent. The estimate obtained for the radius of the plasmoid is in agreement with the results of Section 1. Thus, it has been shown that nanosize plasmoids, described within the framework of the proposed model, can form a composite object due to the quantum exchange interaction between ions.

## 4. APPLICATION OF OBTAINED RESULTS

We advance the hypothesis that quantum nonlinear oscillations of plasma can lie at the basis of a theoretical model of stable, spherically symmetric natural plasma objects [10]. These luminous formations appear mainly during a thunderstorm and their lifetime can range up to several minutes. There are a number of models of the given natural phenomenon, some of which are described in [10]. However, none of the proposed theoretical models explains all of the observed properties of such plasmoids. Thus, the existence of such objects, as before, remains a mystery for modern physics.

Such natural plasma structures probably have an electromagnetic origin. However, as was asserted in [6], the application of classical electrodynamics to describe such plasma objects runs up against insurmountable difficulties.

In [1–4], a quantum approach was developed to describe stable, spherically symmetric plasma structures. These papers describe the formation of bound states of charged particles inside a plasmoid and also the possible creation of a composite plasma structure. Some observable properties of natural plasma objects were also predicted within the framework of the proposed model.

The significance of the obtained theoretical results for the analysis of experiments in which luminous plasma structures were formed as a result of electrical discharges in water (see, for example, [11]), was noted in [1–4]. Application of the developed model to describe nanosize plasmoids recently obtained in a laboratory [12] has also been discussed.

It is interesting to note that one of the observations of an atmospheric plasmoid was recently analyzed in [13]. It was reported that a plasmoid, having a small central region with a size not exceeding 1 mm, melted glass. This observation is in agreement with the predictions of our model. In addition, the high temperature of the plasma at the center of the object, needed to melt glass, is an indication that the given structure does not get its energy from chemical reactions.

Models of natural plasma structures in which microdose nuclear reactions can take place were proposed in [14]. Besides the case mentioned in [13], these models can explain some of the observational data on the very high-energy atmospheric plasma structures described in [15]. In this situation, such atmospheric plasmoids can serve as an alternative energy source, provided they can be successfully reproduced in a laboratory.

The work of one of the authors (M. S. D.) was supported in part by the Program to Enhance the Competitiveness of Tomsk State University Among the World's Leading Scientific and Educational Centers, and also by the Russian Foundation for Basic Research (Grant No. 15-02-00293).

**REFERENCES**

1. M. Dvornikov and S. Dvornikov, in: Advances in Plasma Physics Research, F. Gerard, ed., Nova Science Publishers, New York (2006), Vol. 5, pp. 197 – 212 [physics/0306157].
2. M. Dvornikov, Proc. R. Soc. A, **468**, 415 (2012) [arXiv:1102.0944].
3. M. Dvornikov, J. Plasma Phys., **81**, 905810327 (2015) [arXiv:1311.6875]; J. Phys. A, **46**, 045501 (2013) [arXiv:1208.2208].
4. M. Dvornikov, J. Atm. Sol.-Terr. Phys., **89**, 62 (2012) [arXiv:1112.0239].
5. L. S. Kuz'menkov and S. G. Maksimov, Theor. Math. Phys., **118**, 227 (1999); G. Manfredi and F. Haas, Phys. Rev. B, **64**, 075316 (2001).
6. A. A. Vlasov and M. A. Yakovlev, Theor. Math. Phys., **34**, 124 (1999).

7. T. Neugebauer, *Z. Phys.*, **58**, 474 (1937); G. C. Dijkhuis, Nature, **284**, 150 (1980); B. A. Veklenko, Plasma Phys. Reports, **38**, 513 (2012).
8. P. K. Shukla and A. A. Mamun, Introduction to Dusty Plasma Physics, IOP Publishing, Bristol (2002).
9. A. V. Kulakov and A. A. Rumyantsev, Dokl. Akad. Nauk SSSR, **320**, 1103 (1991).
10. R. K. Doe, in: Forces of Nature and Cultural Responses, K. Pfeifer and N. Pfeifer, eds., Springer Verlag (Berlin), 2013, p. 7.
11. G. D. Shabanov, Tech. Phys. Lett., **28**, 164 (2002); A. Versteeg *et al*., Plasma Sources Sci. Tech., **17**, 024014 (2008).
12. J. B. A. Mitchell *et al.,* Phys. Rev. Lett., **100**, 065001 (2008); T. Ito and M. A. Cappelli, AIP Adv., **2**, 012126 (2012).
13. V. L. Bychkov *et al*., J. Atmos. Sol.-Terr. Phys., **150**, 69 (2016).
14. M. D. Altschuler *et al*., Nature, **228**, 545 (1970); Yu. L. Ratis, Phys. Part. Nucl. Lett., **2**, 374 (2005).
15. V. L. Bychkov *et al*., in: The Atmosphere and Ionosphere: Dynamics, Processing and Monitoring, V. L. Bychkov, G. B. Golubkov, and A. I. Nikitin, eds., Springer Verlag, Dordrecht (2010), pp. 201 - 373.